# Nanothermal Characterization of Amorphous and Crystalline Phases in Chalcogenide Thin Films with Scanning Thermal Microscopy

J. L. Bosse,[1] M. Timofeeva,[2] P. D. Tovee,[3] B. J. Robinson,[3] B. D. Huey,[1] O. V. Kolosov[3,a]

[1]Department of Materials Science & Engineering, University of Connecticut, Storrs, CT 06269-3136
[2]St. Petersburg Academic University, Chlopina 8/3, 194021, St. Petersburg, Russia
[3]Department of Physics, Lancaster University, Lancaster, United Kingdom, LA1 4YB

[a] Electronic mail: o.kolosov@lancaster.ac.uk

**Abstract**

*The thermal properties of amorphous and crystalline phases in chalcogenide phase change materials (PCM) play a key role in device performance for non-volatile random-access memory. Here, we report the nanothermal morphology of amorphous and crystalline phases in laser pulsed GeTe and $Ge_2Sb_2Te_5$ thin films by scanning thermal microscopy (SThM). By performing SThM measurements and applying quantitative physical models to two film thicknesses, the PCM thermal conductivities and thermal boundary conductances between the PCM and SThM probe are independently estimated for the amorphous and crystalline phase of each stoichiometry.*

**Introduction**

Phase change materials (PCM) have been in the focus of research interest for the last decade as a candidate for non-volatile memories such as flash memory and dynamic random access memory as they can combine high read/write speeds, excellent data retention, and low switching power.[1] Phase change memory is based on reversible switching between amorphous and crystalline states,[2] producing remarkable reflectivity contrast for optical devices,[3,4] and electrical conductivity modulation for solid state devices.[5,6] Finding stoichiometries that promote a fast crystallization time, lower threshold switching voltage/current between states, and improved high-cycle reliability are of particular interest.[7] Although various scanning probe microscopy (SPM) techniques have been employed to study these materials by electrical[1,8-11] and nanomechanical[12,13] means, there lacks a quantitative, non-destructive characterization method to investigate local nanoscale thermal properties of PCM that is a critical factor for their switching energy and read/write dynamics. Several methods are currently employed to study thermal properties, such as Raman spectroscopy and IR spectroscopy, however, these have a spatial resolution limited to the micron scale.[14,15] Scanning Thermal Microscopy (SThM),[16] on the other hand, would be ideal for quantitatively measuring and mapping local thermal properties, with the added potential capability of directly reading and writing 'bits' of data (phase changed regions) with spatial resolution down to the nanometer scale.[17,18]

In the present work, we demonstrate a SThM approach to study the thermal properties of amorphous and crystalline phases of commercially viable PCM stoichiometries, $Ge_2Sb_2Te_5$ (a-GST/c-GST) and GeTe (a-GT/c-GT). These are selected as they demonstrate nucleation and growth dominated crystallization behavior, respectively.[19] The thermal responses for the amorphous and crystalline phases are modeled, with thermal conductivities compared with a range of previously reported values. This work is of particular interest to research efforts on determining the phase switching thresholds for phase change materials as a function of varying experimental parameters, such as composition gradients, sample thickness, applied voltage, or power.

**Materials and Methods**

*Sample fabrication and SThM experimental setup*

The phase change materials considered here are the most common commercially viable stoichiometries, GST and GT, which demonstrate nucleation and growth dominated crystallization behavior, respectively. Two different samples were studied for each material, consisting of sputtered

(Moorfield MiniLab 25) 100 and 200 nm thick films of amorphous phase change material on glass coverslips (substrate). In between each phase change film and glass substrate, a 10 nm layer of Ti was sputtered to promote adequate bonding, an order of magnitude thinner than the PCM film, to minimize its influence on measured thermal properties.

Following specimen fabrication, each film was mounted onto a motorized XYZ stage and illuminated with a focused 514 nm wavelength Ar ion laser of varying power from 3 to 4 mW on the sample (Spectra Physics). The specimens were programmatically translated with a step motor controller (Honda Electronics) at 50 μm per second to create crystalline lines in the amorphous films with a consistent heating per unit area. SThM images of various sizes were acquired on the amorphous and crystalline phases of both film thicknesses allowing to investigate the nanoscale thermal properties and corresponding morphology. For the quantitative evaluation of thermal properties, force-distance curves (0.1 Hz ramp rate) were obtained while simultaneously acquiring the thermal signal during approach and retract to determine the temperature drop upon contact with the specimen. Such approach and retract profiles were collected on the crystalline and amorphous regions for both film thicknesses.

All measurements were acquired in ambient temperature and humidity with a Bruker Multi Mode Scanning Probe Microscope with Nanoscope III controller. Thermal transport measurements were performed using SThM probes (Kelvin Nanotechnology, KNT-SThM-01a, 0.3 N/m spring constant, <100 nm tip radius), that were thermally calibrated to relate the probe resistances to probe temperature.[20]

The calibrations were done on a Peltier hot/cold plate (Torrey Pines Scientific, Echo Therm IC20) by incrementing the temperature and recording the probe resistance using ratiometric approach (Agilent

34401A digital multimeter).[21] This was performed externally to the SPM to prevent additional heating from the SPM deflection laser. At first, a DC voltage (sufficiently small to exclude self-heating) was applied to the probe using an arbitrary waveform generator (Agilent 33220A) linking the probe resistance to the ambient temperature. Next, the applied voltage of the probe was increased ensuring controlled self-heating, while recording the probe resistance (Figure S1 in supplemental information) therefore determining the temperature of the probe self-heating as a function of excitation voltage.

The SThM electrical measurements were performed by sensor heating[22,23] with an AC-DC bridge configuration, presented elsewhere in detail by Tovee et al.[20] As expected, the SPM laser illumination for measuring deflection heated the probe by 10°C, therefore adding to the Joule heating of the probe. All measurements were performed with a set-force below 15 nN during imaging to protect the tip and sample from damage to either structure.

For quantification of thermal properties of the phase change specimens, the equivalent thermal resistance between the probe and its surroundings, $R_T$, is considered according to previous models (Figure S2 in supplemental information) as defined by the following equation:[16,20]

$$R_T = \frac{T_H - T_0}{Q_h} \qquad (1.1)$$

where $T_H$ and $T_0$ is the heater and ambient temperature, respectively, and $Q_h$ is the heat generated by the heating element. During the specimen fabrication stage of the experimental process, it is very important to select a sample geometry and substrate that promote optimal contrast for SThM measurements. It is well known from previous experimental data[24,25] that one of the most important factors to consider is the tip/sample thermal boundary conductance (TBC), also known as Kapitza conductance ($\sigma_{ts}$, i.e. the reciprocal of $1/R_{ts}$).[26-29] The SThM response will strongly depend on the tip-sample junction PCM as well as the tip-heater thermal conductance. In preliminary experiments, a 50 nm thin film of phase change material was sputtered onto a substrate of doped Si. However, the Si substrate, having a very high thermal conductivity, negatively impacted the results of the PCM thin film by masking any local conductivity variations between the tip and sample. By selecting a much thicker phase change film (100-200 nm) and substrate with significantly lower thermal conductivity (soda lime glass), the heat transport in the PCM film dominated the measurements, producing notably better results (i.e. stronger SThM sensitivity to the varying properties of the phase change material and resulting resolution of nanoscale features). Additionally, by performing SThM measurements on two different film thicknesses, i.e. 100 and 200 nm films, and assuming a uniform TBC regardless of thickness (a reasonable approximation given that the mean free path of the heat carriers (phonons) in PCM is much shorter than film thicknesses involved),[30] the true sample thermal conductivity may be extracted from the experimental SThM data.

*Multi-scale modeling*

A three dimensional finite element analysis (FEA) was performed using commercial software (COMSOL Multiphysics, Joule Heating module) in order to determine the influence of cantilever/sample geometry and sample materials properties on the SThM experimental results. By isolating and understanding the interplay of these factors, the thermal conductivities of the amorphous and crystalline phases can be estimated.

The FEA model is based on the experimental setup as described, with a SThM cantilever, GST or GeTe thin film, soda-lime-silica glass substrate, and thin Ti interlayer between the PCM and substrate. The proportions and materials used for the modeled SThM cantilever were similar to those implemented in the experiments, Figure 1(a), with 250 nm Au pads and 150 nm Pd resistors micro-patterned on a commercial $Si_3N_4$ cantilever base.[20] The modeled PCM samples consist of a 2 μm x 8 µm crystalline phase positioned between two 8 µm x 8 µm amorphous phases, with a thickness equal to either 100 or 200 nm (based on the two SThM experimental cases considered). The cantilever and sample were placed in air environment, and the temperature profile of the entire 3-Dimensional system was calculated, Figure 1(b). The thermal conductivities for all materials used in the 3D model are presented in Table 1. Note that the thermal conductivities of the sputtered Au pads and $Si_3N_4$ cantilever base, with effective values of 170 and 4.5 W/m-K, respectively, are determined by matching the heat-temperature balance and conductance values of the SThM probe in air (within 0.25 to 0.50 K at 293 and 353 K) with experimental data for both hot plate and self-heating calibration measurements, while accounting for the electrical circuit of the probe containing two 100 Ω resistors in series with the heater.

As discussed previously, the tip-sample TBC ($\sigma_{ts}$) is important for accurately accounting for the thermal resistance of the probe-sample junction, and therefore for determining the thermal conductivity of the amorphous and crystalline phases for each stoichiometry we studied. This may be presented as:

$$\sigma_{ts} = \rho_c \, \pi \, r_{ts}^2 \qquad (1.2)$$

where $\rho_c$ and $r_{ts}$ are the conductance and effective interface radius of the contact between the tip and sample, respectively. To incorporate the TBC in the FEA simulation, we include a thin resistive layer between the tip apex and the sample represented by a cylinder with height ($h$) much smaller than the contact diameter ($2\,r_{ts}$). The thermal conductivity of the TBC is then calculated as:

$$\sigma_{ts} = h * \rho_c \qquad (1.3)$$

The thermal transport in most regions of the modeled SThM system can then be handled by standard heat transfer equations. The characteristic dimensions of the structural elements are on the order of 50 nm to micrometres, which in all cases are higher than the mean free paths in the various materials components, allowing use of the diffusive heat transfer equation:[31]

$$\rho C_p \frac{\partial T}{\partial t} = k \Delta T + Q \qquad (1.4)$$

where $\rho$ is the density of the material, $C_p$ is the heat capacity at constant pressure, $k$ is the media thermal conductivity, and $Q$ is the heat source. The temperature distribution is assumed to be time independent due to the slow ramp rate of the force-distance curves, so the left hand side of eq. 1.4 equates to zero. By solving eq. 1.4 for all structural parts of the system (Figure S2 in Supplemental Information) and with the proper boundary conditions, we then obtain the modeled temperature distribution. The thermal boundary conditions were set such that the temperature of the surrounding environment as well as the initial temperature of all domains was 293 K. A fixed electrical potential difference is applied across Pd resistors at the probe apex as identified in Figure 1(a) (the only domain in the model to include an electrical component) to induce local Joule heating reflecting experimental conditions. Finally, the thermal discontinuity experienced by the probe when brought into contact or out of contact was calculated and compared to that of corresponding experimental data. By adjusting the thermal properties of the modeled amorphous and crystalline phases to match the SThM experimental results, the measured amorphous and crystalline PCM thermal properties are estimated.

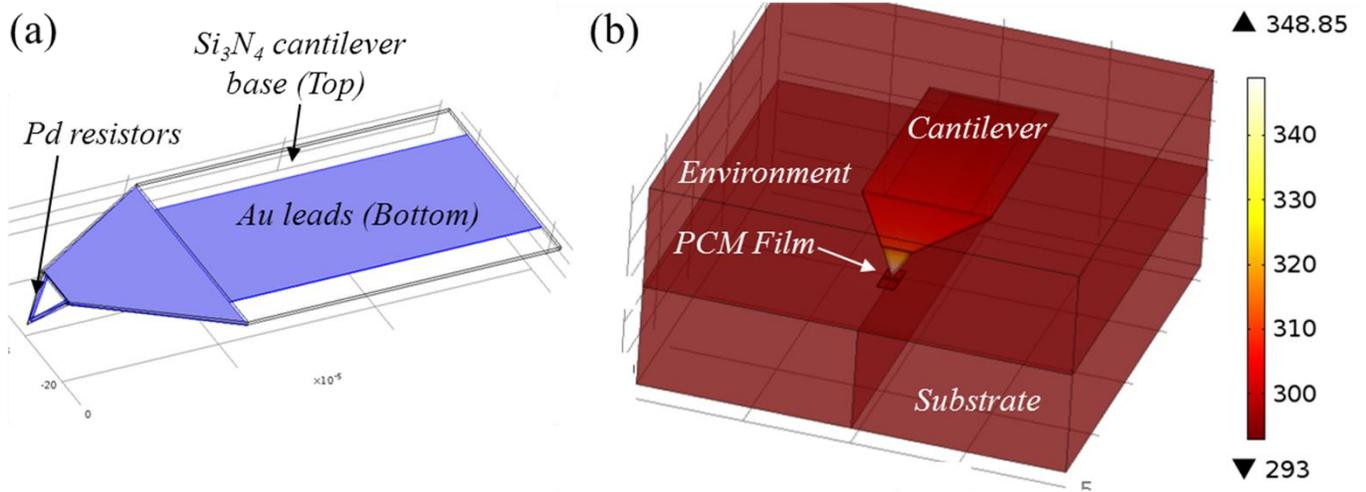

**Figure 1:** The design of the SThM probe with Si$_3$N$_4$ cantilever base, Au pads, and Pd resistors (a) reflected in the simulation. The model system comprises a cantilever approaching a PCM film on a soda lime glass substrate, and the ambient air environment (b).

**Table I:** Thermal conductivities (W/m-K) for all materials used in the FEA model. *Note that effective values are used for Au and Si$_3$N$_4$ thin films to match the experimentally measured probe thermal and electrical resistances for the hot plate and self-heating calibration measurements.

| Pd | Soda-lime glass | Air | Ti | Au | Si$_3$N$_4$ |
|---|---|---|---|---|---|
| 71[32] | 1.05[33] | 0.02[34] | 21.9[32] | 170* | 4.5* |

**Results and Discussion**

Figure 2(a,c) presents representative topography (left) and corresponding SThM (right) images for the 200 nm GT specimen with 10 and 2.5 µm scan sizes, respectively. Figure 2(b,d) presents similar results for the 200 nm GST specimen, but with 8 and 2.5 µm scan sizes, respectively. The SThM images display the temperature of the SThM sensor, henceforth labeled as 'thermal images' with constant power applied to the probe.

Topographically, the depressions running down the centers of the height images correspond to the crystalline phases nucleated in the surrounding amorphous film by the pulsed laser as it traversed the film. Such a specific volume reduction between amorphous and crystalline phases is expected, and is typically 5% for these stoichiometries.[35]

For the SThM images, the thermal response is uniformly darker (decreased contrast) for the crystalline phase compared to the surrounding amorphous film, indicating a combined tip-sample and sample spreading thermal resistance, $R_{ts} + R_s$, which is lower than for the amorphous regions.

There are two noteworthy aspects related to the morphology at the boundary between the amorphous and crystalline phases. The higher magnification SThM images in Figures 2(c,d) identify that the boundary is sharper for GT versus GST. Figure 2(c) reveals a 30 to 50 nm transition between the crystalline and amorphous regions for GT. For GST, on the other hand, the gradient in thermal properties from crystalline to amorphous regions occurs over 80 to 440 nm, Figure 2(d). Furthermore, the crystal/amorphous boundary represents a relatively straight line for the GST film, while for GT it has clear deviations from such line. While the line undulation may be expected due to the discrete motion of the step motor, the fact that it is more prominent for the GT film may relate to the growth dominated crystallization behavior for GT as compared to GST, causing more variability in GT phase boundaries once nucleation sites become activated.

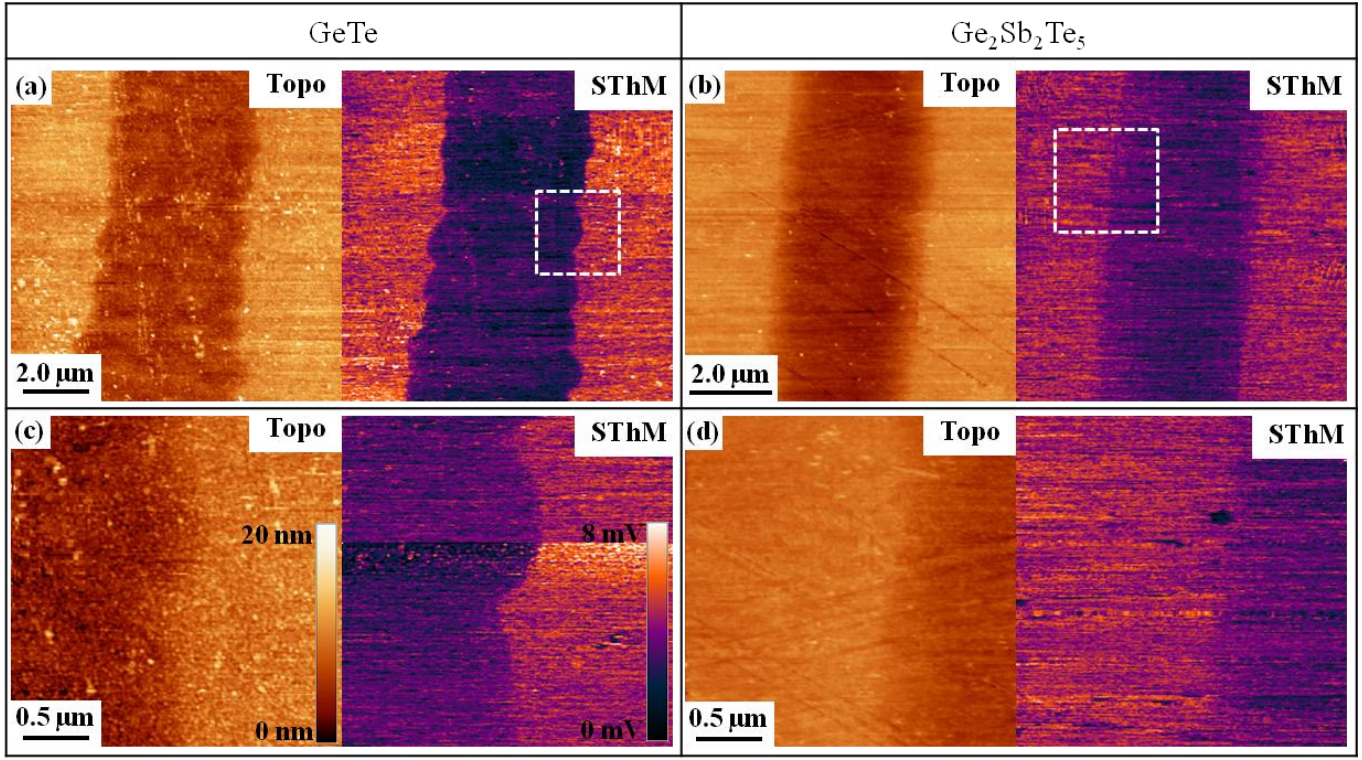

**Figure 2:** Topography (left) and SThM (right) images with 10 μm (a) and 8 μm (b) scan sizes, respectively, revealing a crystalline line written into 200 nm GT and GST amorphous thin films by a scanned, pulsed laser. The 2.5 μm images (bottom row) for GT (c) and GST (d) are taken from the spatial locations marked by the insets in (a,b).

To better quantify the thermal properties of the crystalline and amorphous regions of each film, conventional 'force-distance' curves were acquired with the SThM tip by approaching until contacting the surface, then retracting until separation, recording the piezo displacement, lever deflection (normal force), and thermal signal throughout. Figure 3(a,b) presents such force-distance curves for 200 nm crystalline and amorphous GST films, respectively, with the tip approaching from the left, snapping to contact leading to a slight decrease in deflection, then linearly deflecting positively as the displacement increases further, indicating that the SThM lever is highly compliant compared to the sample. Figure 3(c,d) presents the simultaneously acquired thermal signals, with approach (dashed) and retraction (solid) directions also identified as shown. While approaching the sample, the thermal signal decreases linearly until the point of tip/sample contact (compare with the snap-in displacements from Figures 3(a,b)), at which point the signal abruptly decreases due to the added tip-sample thermal conductance. During tip retraction, adhesion forces maintain contact until pull-off occurs as is typical for AFM-based measurements in ambient conditions. The thermal signal again changes sharply, now due to loss of contact, after which the thermal response matches the previous, non-contact values.

When comparing the crystalline (Figure 3(c)) to amorphous (Figure 3(d)) thermal approach curves, the thermal drop is notably stronger for the crystalline phase, consistent with the SThM imaging performed in Figure 2 where the crystalline regions exhibit lower signals. To quantify this parameter more thoroughly, such sharp drops and the subsequent rise in the thermal response for approach and retract, respectively, were averaged for several groups of successive force-distance curves (N=3) and analyzed for each stoichiometry, specimen thickness, and phase. The approach portion of these experimental results was then compared with thermal modeling for equivalent conditions. It is worth noting that the retract curves could have also been used for comparisons to thermal modeling, as experimentally they display similar trends as observed in Figure 3. However, the magnitudes of the thermal jumps are generally less reliable since retraction curves also depend on adhesion effects during tip/sample pull-off. An increase in adhesion would thus produce a larger pull-off displacement (~75 vs. ~40 nm for crystalline and amorphous GST, respectively, in Figure 3), and hence a greater pull-off deflection (~150 vs. ~60 nm), distorting interpretation of the corresponding thermal jump as if a higher thermal conductivity were encountered. The snap-to-contact displacement, and deflection, for approach curves are susceptible to adhesion to much smaller degree with this nearly uniform change in lever deflection (~20 nm). Therefore, any error caused by such adhesion-based artifacts (if present) is minimized for approach curves that are therefore preferred for the SThM quantitative measurements.

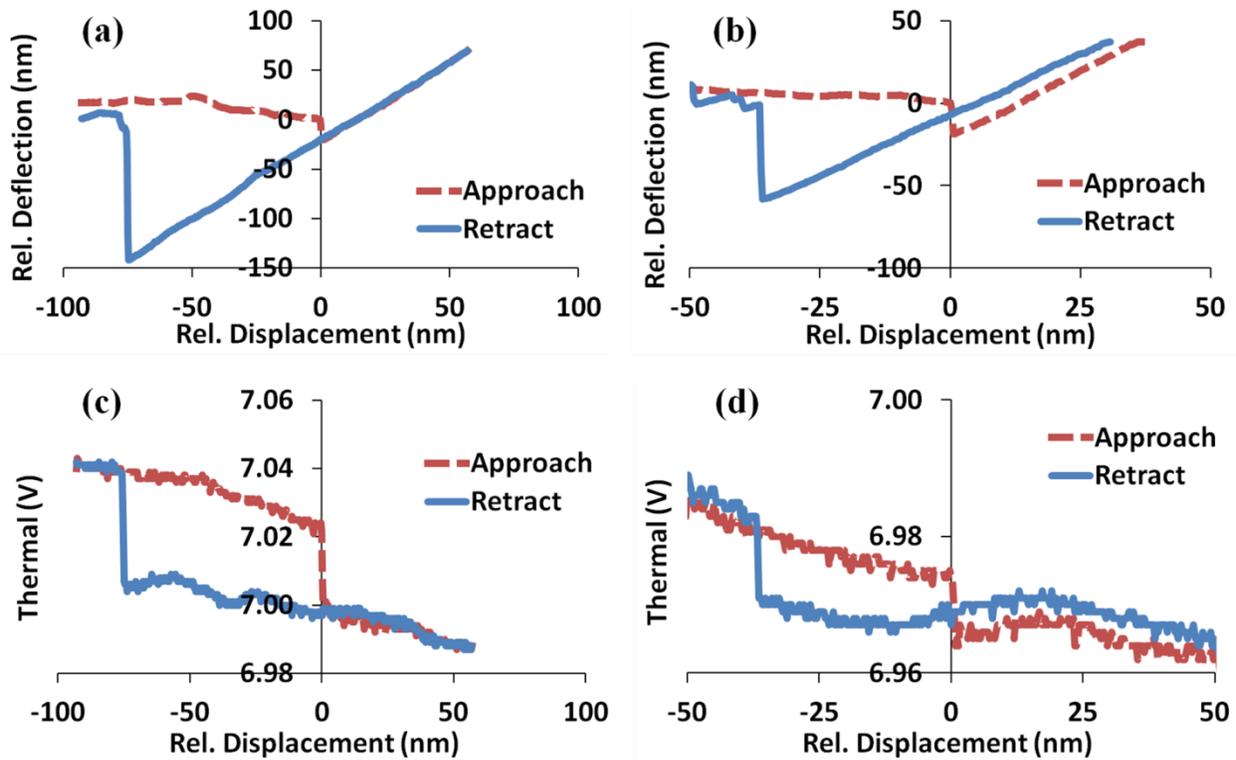

**Figure 3:** The typical approach and retract SThM curves for PCM materials, with simultaneously recorded relative cantilever deflection (a,b) and thermal (c,d) signal as a function of relative displacement for contact with crystalline (a,c) and amorphous (b,d) GST phases.

The observed thermal "drops" upon contact with the crystalline phases are consistently larger regardless of film thickness, and for both GST and GT (not shown for brevity). However, the contrast between the crystalline and amorphous phase is stronger for thicker PCM films, as anticipated due to the larger contribution of the film with respect to the underlying glass substrate. Since tip-sample TBC $R_{ts}$ is identical for both measurements but the thermal resistance of the film $R_s$ differs with the film thickness, the tip-sample contact resistance can be extracted with appropriate models. The TBC is determined using the acoustic mismatch model (AMM).[27] It is then considered as an equivalent cylinder representing the tip/sample contact area (see Methods), and the modeled thermal drops are calibrated to match the experimental values. Finally, the influence of the TBC is removed to determine the thermal conductivity of each phase and material.

The temperature distribution of the modeled SThM system is presented for the SThM probe out of contact (Figure 4(a)) and in contact (Figure 4(b)) with c-GST, as well as out of contact (Figure 4(c)) and in contact (Figure 4(d)) with a-GST. The model accounts for the substrate, underlying adhesion layer, chalcogenide film, environment, probe geometry near the apex, and distinct probe materials including a primary silicon nitride tip as well as the resistive heating elements.

For contact with the crystalline GST film, heat is conducted easily from the probe in the plane of the film, and through the glass substrate. This predicts the largest temperature drop of the probe, as measured experimentally. For contact with the amorphous GST film, on the other hand, the higher thermal resistance limits heat dissipation in-plane as well as into the glass substrate, retaining more heat locally. As a result, a weaker thermal drop is predicted, and experimentally measured. When out of contact, the highest temperature of the probe is observed, with minimal heat loss to the PCM and underlying glass substrate as expected. Nevertheless, for near-contact conditions as modeled (50 nm separation), the a-GST (Figure 4(c)) is noticeably hotter than the c-GST out of contact (Figure 4(a)). Equivalent in- and out-of-contact thermal distributions were prepared for c- and a- phase GT, but again are omitted as they follow a similar trend.

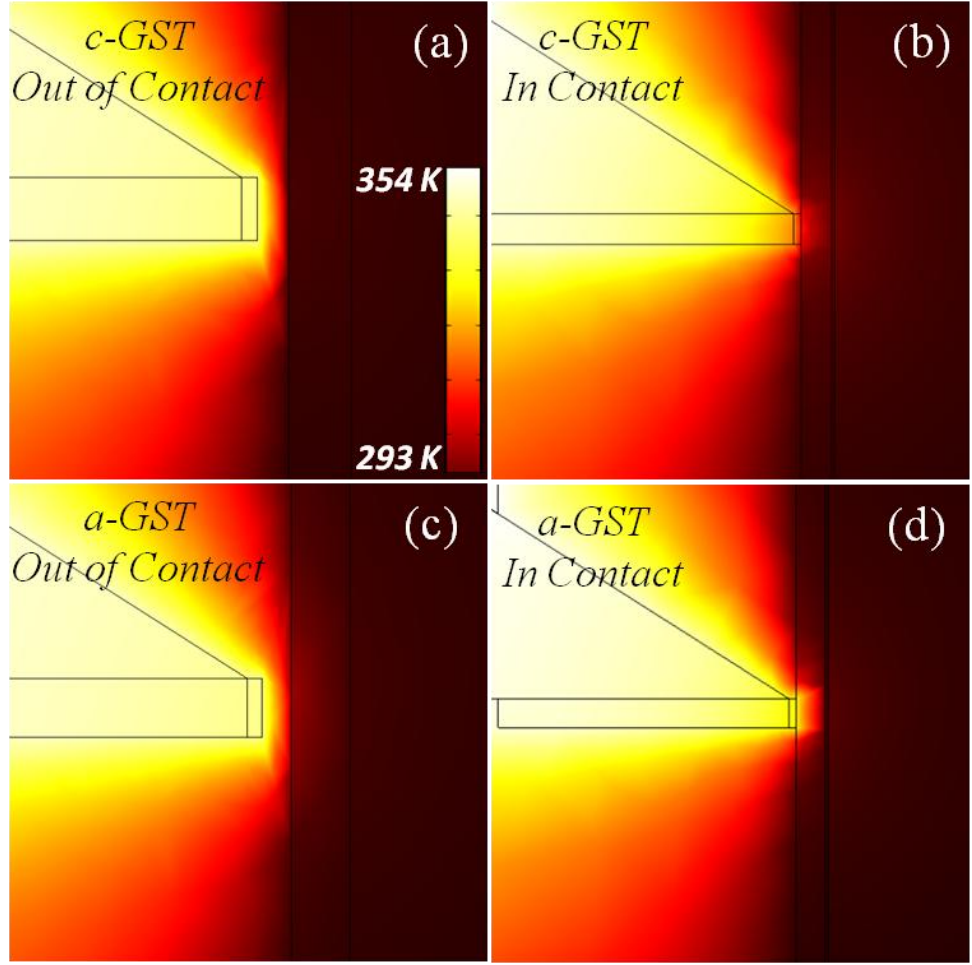

**Figure 4: Cross-section view of the simulated temperature distribution between the SThM probe and sample. (a) Out-of-contact and (b) in-contact data for 100 nm c-GST vs out-of-contact (c) and in-contact (d) of 100 nm a-GST film. The out of contact tip/sample distance is 50 nm, and the temperature scale bar applies to all cases. Although not fully visible in (a) and (c), the 10 nm Ti layer is present and incorporated into the temperature distribution model.**

The thermal conductivities of a-GST/c-GST (Figure 5(a)) and a-GT/c-GT (Figure 5(b)) thin films are finally calculated by iteratively fitting the model to the experimentally acquired thermal drops. As presented in Table II, the resulting thermal conductivities for a-GST and c-GST are 0.30 and 1.95 Wm$^{-1}$K$^{-1}$, respectively, while they are 0.20 and 1.60 Wm$^{-1}$K$^{-1}$ for a-GT and c-GT. These locally measured thermal conductivities for a-GST and c-GST are within the range of values determined by previous studies using more macroscopic methods, 0.19 to 0.33 Wm$^{-1}$K$^{-1}$ [27,36,37] and 1.1 to 2.0 Wm$^{-1}$K$^{-1}$,[36,37] respectively. The particularly high a-GST value may be explained by considering film preparation, where elevated temperatures during sputtering could result in the presence of a small fraction of nucleated crystalline phase as observed in separate mechanical studies[38] and hence a higher effective thermal conductivity. Additionally, as the experimental a-GST phase was placed between two c-GST reference lines, that may also somewhat contribute to increased heat conduction and therefore result in a higher observed thermal conductivity. Finally, standard deviation error bars reveal a higher uncertainty for the crystalline phase of each stoichiometry. This results from a stronger variation in the experimentally measured thermal "jumps" for the crystalline regions. This can be linked to variations in the local crystallite orientations under the SThM probe and hence a wider range of directionally dependent thermal properties. The resulting a-GT and c-GT thermal conductivity values are considerably lower than those previously reported,[39] 2.3 and 5.7 $Wm^{-1}K^{-1}$ for a- and c-GT, respectively. However, the discrepancy in the values may be explained by the contrasting measurement methods. For example, the thermal conductivity measurements on a- and c-GT by Nath and Chopra[39,40] were acquired on a 900 nm film at steady-state, by an in-plane thermal gradient over a 4.0 x 0.5 cm length scale, clearly demonstrating a convergence with bulk values. Here, the thermal gradient was applied normal to the thin film surface, with heat flow considered over an area 6 orders of magnitude less.

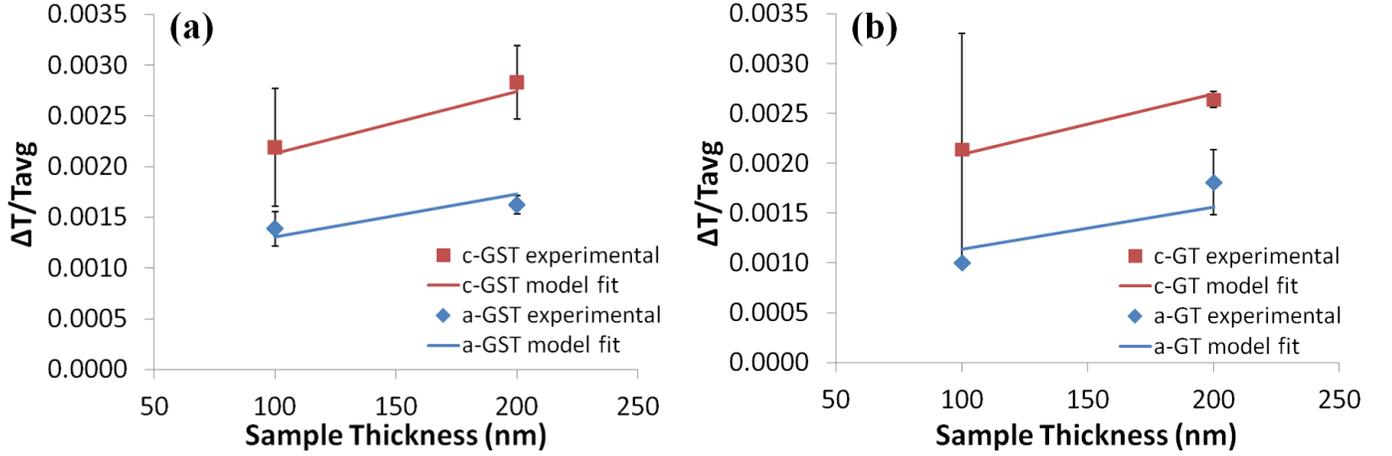

**Figure 5:** Normalized thermal drop (ΔT/Tavg) versus sample thickness for amorphous and crystalline GST phases, including experimental data (with standard deviation error bars, N=3) and a model fit (a). Similar data is presented for amorphous and crystalline GT (b).

**Table II:** Thermal conductivities (W/m-K) for amorphous and crystalline phases of GST and GT, acquired by fitting the simulated temperature profile of the probe to those measured experimentally with force-displacement curves.

| Phase | a-$Ge_2Sb_2Te_5$ | c-$Ge_2Sb_2Te_5$ | a-GeTe | c-GeTe |
|---|---|---|---|---|
| Thermal conductivity [W/m-K] | 0.30 | 1.95 | 0.20 | 1.60 |

The TBC between GST films and substrates of different materials (C, Ti, TiN) has been calculated elsewhere using the acoustic mismatch model (AMM).[27] However, thermal time-domain thermoreflectance (TDTR) data reveals approximately one order of magnitude lower conductance values due to interfacial effects such as grain boundaries, impurities, and surface defects.[41] For example, AMM values range from $5.0 \times 10^8$ to $3.3 \times 10^{10}$ Wm$^{-2}$K$^{-1}$ and $5.3 \times 10^8$ to $1.4 \times 10^{10}$ Wm$^{-2}$K$^{-1}$ for a-GST and c-GST, respectively, while TDTR values range from $3.9 \times 10^7$ to $5.6 \times 10^7$ Wm$^{-2}$K$^{-1}$ for c-GST (no data is available for a-GST). The TBC values for a- and c-GST in contact with a $Si_3N_4$ SThM probe as implemented here have not been reported, so values were calculated instead based on the acoustic mismatch and geometry,[26,42] specifically $7.0 \times 10^8$ and $3.8 \times 10^7$ Wm$^{-2}$K$^{-1}$ [27] between a-GST/$Si_3N_4$ or c-GST/$Si_3N_4$ contacts, respectively. TBC values for a- and c-GT in contact with the

Si$_3$N$_4$ probe have also not been explicitly reported, so the a- and c-GST values were applied; a reasonable assumption as the GST/GT Debye temperatures are similar.[43,44]

**Conclusion**

Scanning thermal microscopy (SThM) has been implemented to characterize optically switched chalcogenide phase change materials of GeTe (GT) and Ge$_2$Sb$_2$Te$_5$ (GST). Quantitative physical models together with the experimental results allowed to account for the thermal boundary conductance, and to directly determine both the thermal conductivities of the amorphous and crystalline phases as well as contact thermal resistances. The thermal conductivities for amorphous and crystalline GST are 0.30 and 1.95 Wm$^{-1}$K$^{-1}$, respectively. The thermal conductivities for amorphous and crystalline GT are 0.20 and 1.60 Wm$^{-1}$K$^{-1}$, respectively. The reported approach has been demonstrated as an effective tool for measuring thermal properties of nanoscale phase change materials, while distinguishing thermal contrast of distinct phases down to 50 nm. SThM provides an alternative characterization method to IR imaging or Raman micro-spectroscopy, and is applicable for the characterization of other thin film materials with similar low thermal conductivities.

**Acknowledgements**

JB and BDH recognize DOE, Basic Energy Sciences, Electron and Scanning Probe Microscopies, grant DE-SC0005037 for support. OVK acknowledge support from the EPSRC grants EP/G06556X/1, EP/K023373/1 and EU grants QUANTIHEAT and FUNPROB.

# Supplemental Information

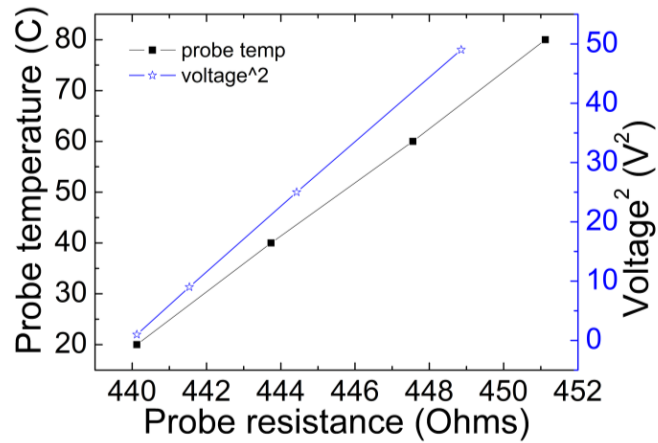

**Figure S1: SThM probe calibration at 1 $V_{DC}$.** The left axis presents the probe temperature as a function of probe resistance, while the right axis presents self-heating of the probe. The quadratic scale indicates a linear increase in probe resistance with Joule heating power (~$V^2$).

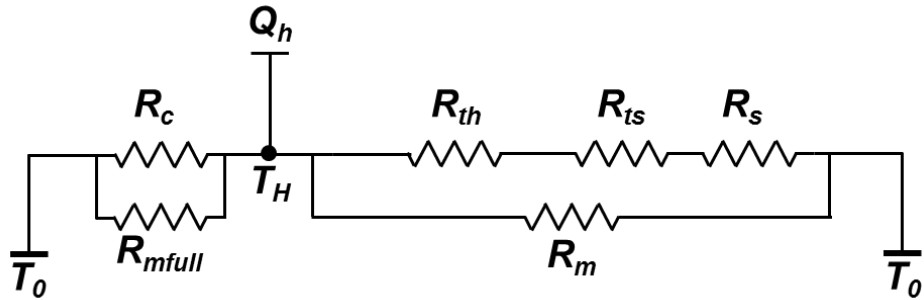

**Figure S2:** The equivalent thermal resistance between the probe and its surroundings are considered according to previous models.[20] The heat generated by the heating element, $Q_h$, is transferred through the environment surrounding the cantilever holder, via thermal resistance $R_{mfull}$, and through the cantilever base itself, via thermal resistance $R_c$. At the tip apex, the heat is transferred to the environment, $R_m$, through the tip and heater, $R_{th}$, and while in contact with the sample, the tip-sample contact resistance, $R_{ts}$ and sample resistance, $R_s$. The heat generated will create a thermal gradient between the heater, $T_H$, and ambient environment, $T_0$.